\documentclass[10pt,letterpaper,twocolumn,aps,showpacs,preprintnumbers,amsmath,amssymb,nofootinbib,prl]{revtex4-2}
\usepackage{graphicx}
\usepackage{dcolumn}
\usepackage{mathrsfs}
\usepackage{bm}
\usepackage{lineno}
\usepackage[utf8]{inputenc}
\usepackage[T1]{fontenc}

\usepackage{color}
\usepackage{amsmath}
\usepackage{amssymb}
\usepackage[unicode=true,pdfusetitle, bookmarks=true,bookmarksnumbered=false,bookmarksopen=false,
 breaklinks=false,pdfborder={0 0 0},pdfborderstyle={},backref=false,colorlinks=true]
 {hyperref}
\usepackage{bbm}
\usepackage{ulem}
\usepackage{xcolor}

\begin{document}

\title{Quantum telescopy with realistic anti-correlated optical fields as references} 

\author{Xuan Tang and Z. Y. Ou\texorpdfstring{$^{\dag}$}{}} 

\affiliation{\mbox{Department of Physics, City University of Hong Kong, 83 Tat Chee Avenue, Kowloon, Hong Kong}\\ 
\texorpdfstring{$^{\dag}$}{}jeffou@cityu.edu.hk}

\date{\today}

\begin{abstract}
First proposed by Grottesman et al [Phys. Rev. Lett. 109, 070503 (2012)], the quantum telescopic technique is to use spatially entangled single-photon states as references to modify the traditional Hanbury Brown-Twiss (HBT) intensity stellar interferometric technique. The technique can rival the Michelson stellar interferometry in signal level but is not restricted by the baseline size, so it has the potential to significantly increase the resolution in astronomical imaging in the optical domain. In practice, single-photon states generated in the lab are usually not ideal in that they have finite bandwidth and non-unit efficiency. Furthermore, detectors also have finite response time and celestial light has a bandwidth. All of this will have an impact on the effectiveness of the quantum telescopic technique. In this work, we consider  a spatially entangled singe-photon state generated by beam-splitting a practical anti-bunched photon source from resonance fluorescence, which is a common method of generating single-photon states, and evaluate the signal level in quantum telescopy with inclusion of imperfection of detection system and finite bandwidth of celestial light. Here, given the incoming light, we optimize the parameters of the anti-bunched photon source and the detector system for the best signal-to-noise ratio (SNR), which can match that of the Michelson method in the ideal case but is always smaller in the realistic case.   Alternatively, weak coherent states from heavily attenuated lasers can replace the entangled single-photon state but suffer low signal level, similar to HBT method. We find that the SNR with a realistic anti-bunched source for producing the reference fields can still beat the SNR of the coherent state scheme in the weak incoming field limit. The optimized parameters for the anti-bunched source and detection system will be a guide for future experimental endeavor.

\end{abstract}

\maketitle 

\section{Introduction}

Due to diffraction of waves, the angular resolution of any imaging system is limited to $\lambda/b$ for a radiation field at a wavelength of $\lambda$ and an aperture size of $b$. That is why all telescopes have large size, not only for collecting more light but also for higher resolution. The practical limitation of the size of the telescopes leads to the synthetic aperture technique which combines images from multiple telescopes separated by large distances through the interference technique, as first achieved in Michelson’s stellar interferometry \cite{mich20,mich21} and currently in operation as the Very Large Telescope Interferometer (VLTI) \cite{VLTI}. The radio frequency version (RF) is the very long baseline interferometry technique (VLBI) that combines electronic signals from radio telescopes at locations separated by a large distance \cite{VLBI}. For optical waves, however, since the most common method of detection is intensity measurement, which loses phase information after detection, observation of optical interference requires light to merge within the coherence length \cite{mich20}, so interference occurs before intensity detection. Because of the difficulty in balancing large optical paths within the coherence length, the baseline size in the optical regime is currently limited to a few hundred meters. 

To circumvent this problem, Hanbury-Brown and Twiss (HBT) invented the stellar intensity interferometer \cite{HBT} by measuring the intensity correlation between two spatially separated telescopes (Fig.1(a)) without the need to bring light together, similar to VLBI. However, due to the lack of phase information, the method is only applicable to symmetric objects such as stars to obtain their diameters. Moreover, the intensity correlation method relies on the coincidence of photons between telescopes, but celestial photons are uncorrelated so that the coincidence signal is orders of magnitude smaller than the direct intensity signal of Michelson’s stellar interferometric method.  

    \begin{figure*}[t]
        \includegraphics[width=13cm]{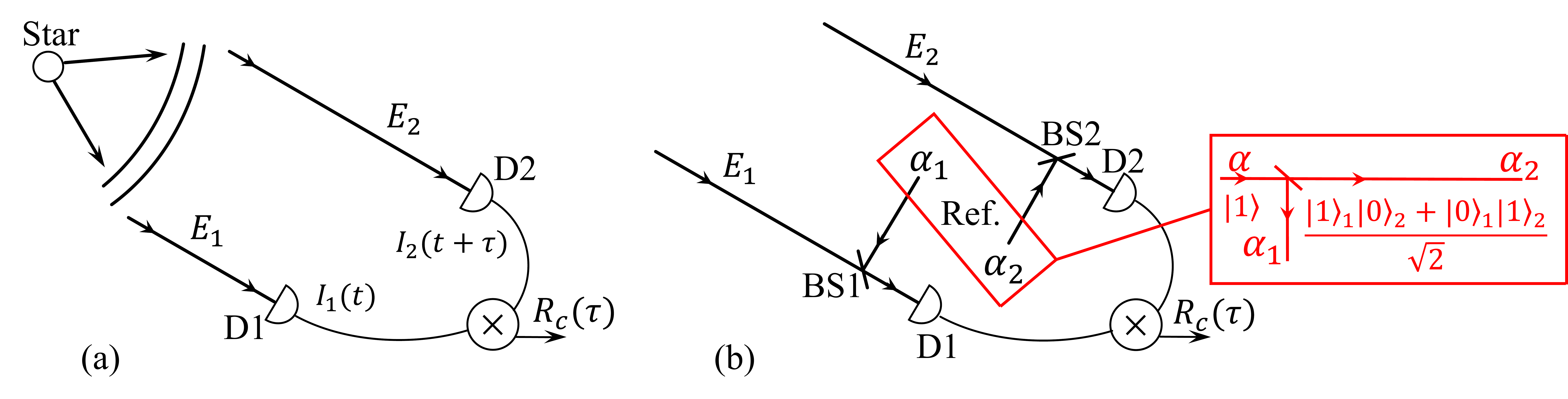}
            \caption{(a) Traditional HBT intensity interferometer; (b) Phase sensitive HBT scheme with coherent reference fields.  Red inset: Creation of coherence between reference fields by beam splitter and use of a spatially entangled single-photon state as the reference fields as proposed by Grottesman et al \cite{gott}. }
            \label{HBT}
    \end{figure*}

In 2012, Grottesman et al (GJC) \cite{gott} proposed that spatially entangled nonlocal single-photon states in the form of $(|1_1,0_2\rangle + e^{i\Delta\phi}|0_1,1_2\rangle/\sqrt{2}$ (Fig.1(b)) can be used as reference fields in a modified HBT interferometer for long-baseline astronomy applications of high-resolution optical imaging, with potential resolution far better than current state-of-the-art radio telescope arrays. The scheme was theoretically proven to have a signal level that matches that of the Michelson method in the weak input field limit \cite{tsa11}. The idea, known as quantum telescopy, was tested in a tabletop setting with heralded pulsed single photons \cite{smi23,th23}, and more recently with nonlocal entangled photon sources supported by quantum repeaters \cite{qr1,qr2}, which predominantly emphasize entanglement distribution in quantum networks. However, these sources are not the ideal single-photon sources required for the quantum telescopy application because they have a finite bandwidth and usually operate in pulsed mode, which cannot temporally match celestial light in CW mode. Passive modification of celestial light or post-selection of photon events can match the temporal modes but at the expense of losses of celestial light.This approach was used by Deng et al. \cite{deng} to demonstrate for the first time the interference between celestial light (Sunlight) and single-photon sources on earth, which is the basis for quantum telescopy.

It is known that under heavy losses, a single-photon state is mixed with vacuum state and can be approximated by a weak coherent state, which then can be a temporary substitute before the technology of high fidelity transmission of entangled single-photon states is mature, as suggested by Grottesman et al \cite{gott}, and the feasibility has recently been demonstrated \cite{ou22,nom23,tang25} as the phase dependent HBT effect. In addition to being readily available with high rate, another advantage of using coherent sources to replace entangled single-photon sources is that their phases can be easily stabilized to local optical clocks without the need for transmission. However, their use as references in quantum telescopy suffers from a low signal level issue, as proven in Ref.\citenum{tsa11}. This is similar to the original HBT intensity interferometer. 

The spatially entangled single-photon state in the GJC scheme can be produced by beam splitting a single-photon state. Current state-of-the-art single-photon sources, with innovative techniques such as microcavity-enhanced Purcell effects, can achieve high emission rates at the GHz-level with an exceptional quality of $g_2\sim 0.02 \ne 0$ \cite{tom21,ding25}. These single-photon sources operating in the CW mode, although non-ideal ($g_2 \ne 0$) and difficult to produce and transmit with high fidelity, can offer an enhancement in signal level, making them a better choice and worthy of further investigation. 

From a practical experimental point of view, in addition to the non-zero $g_2$-value, the current state-of-the-art CW single-photon sources have a finite correlation time due to finite bandwidth. This means that even $g_2$ has a zero value, it is so only within a finite time period of its correlation time. The photon rate is also limited by the correlation time because the separation between photons cannot be smaller than the correlation time.   Moreover, optical detectors have a finite response/resolving time that may limit the observable bandwidth of celestial light as well as the use of the single-photon state with finite bandwidth.

The aim of this work is to consider all these practical constraints in the application of single-photon states as reference fields to quantum telescopy. We will find the scope of the experimental parameters that can optimize the signal-to-noise ratio (SNR). We compare it with the Michelson direct intensity measurement scheme as well as the scheme of using coherent states as reference fields.  We find that for weak incoming fields, the SNR of the scheme with single-photon states can match that of the Michelson scheme only under nearly ideal conditions but is always better than that of the scheme with coherent states. We investigate how to approach the ideal case through quantum state engineering.   

The outline of the paper is as follows: we start in Sect. II by first describing the phase sensitive HBT effect with arbitrary sources as the references. Next in Sect.III, we consider practical constraints of anti-bunched light fields and the finite response time of the detection systems. In Sect.IV, we look into a practical anti-bunched photon source of resonance fluorescence. In Sect.V, we give a formula of the observed signal-to-noise ratio (SNR) as the criterion for comparison. We optimize the SNR in Sect.VI with various source and experimental parameters. In Sect. VII, we consider the effect of losses on SNR. In all of the discussion, we will use the scheme of Michelson stellar interferometry as a reference for comparison. For the HBT scheme, we will compare the results for using either coherent states or anti-bunched photon sources as the references and obtain an enhancement factor for the quantum sources.

\section{Phase dependent HBT effect with anti-bunching light} 
    
\begin{figure}[t]
        \includegraphics[width=8.5cm]{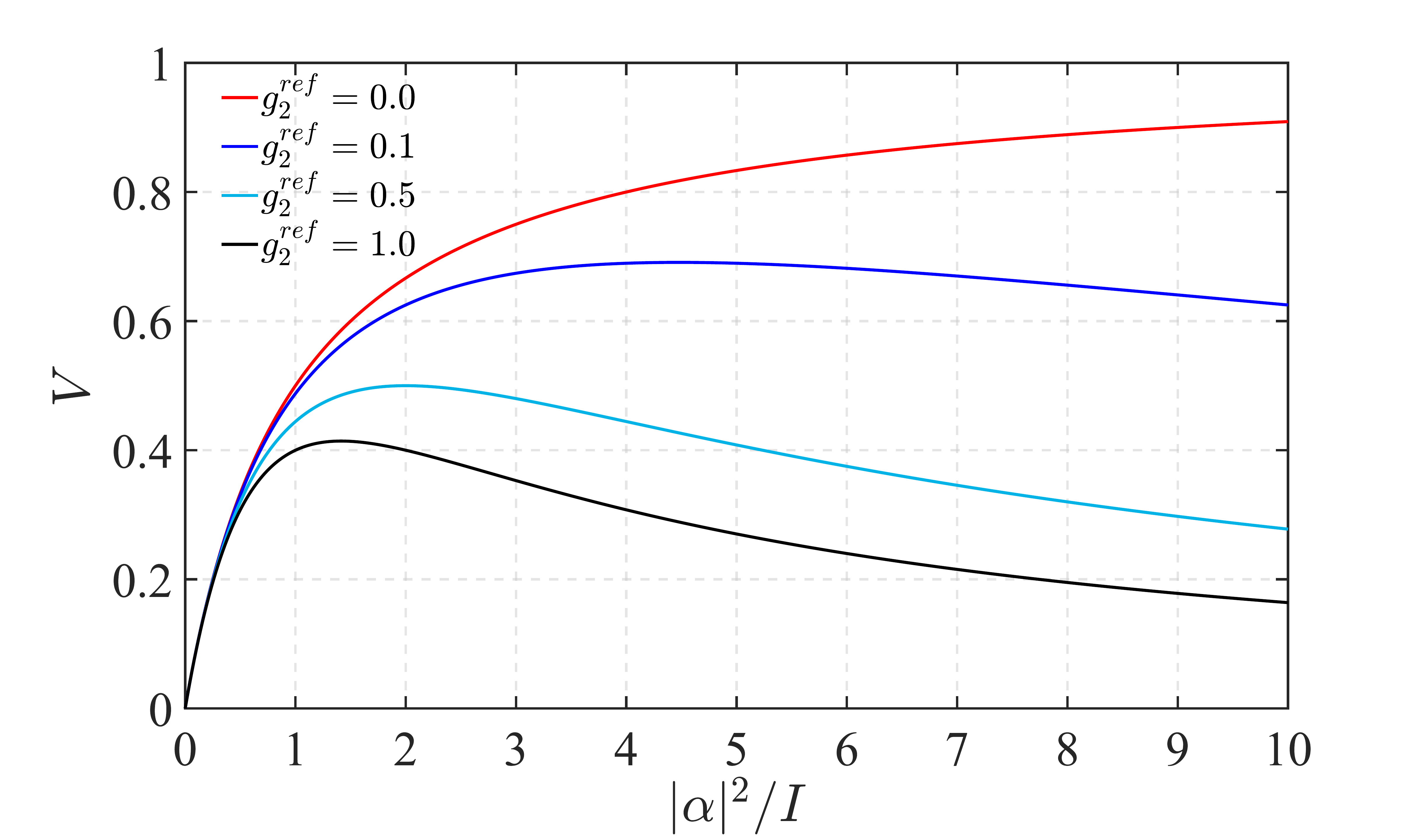}
            \caption{Visibility of phase-dependent HBT effect as a function of $|\alpha|^2/I$ for reference sources with $g_2^{ref}=0$ (red), 0.1 (blue), 0.5 (sky blue), 1.0 (black).} 
            \label{Vi}
    \end{figure}

The phase-dependent Hanbury Brown–Twiss (HBT) effect represents an extension of the traditional phase-insensitive HBT intensity interferometry (Fig.\ref{HBT}(a)), achieved through the introduction of  reference light fields that are mixed with input fields before detection (Fig. \ref{HBT}(b)). To measure the phase coherence between the input fields, the reference fields must also be coherent, which can be done by splitting a common field with a beam splitter  or using the spatially entangled single-photon state proposed by GJC, as shown in the red dashed box of Fig. \ref{HBT}(b). In our previous work \cite{tang25}, an attenuated laser is split to serve as reference sources, enabling reconstruction of the full complex degree of coherence between two optical fields at separated locations. Similarly to the HBT scheme, coincidence measurements between the two detectors are performed to show an interference pattern from which the phase information is extracted. For arbitrary sources as references, it can be shown that the intensity correlation between detector D1 and D2 in Fig.\ref{HBT}(b) is given by \cite{tang25}
    \begin{equation}
        \begin{aligned}\label{corr}
            \Gamma_{2,2}(\tau ) & = I_1I_2g_2(\tau)+{|\alpha _1\alpha _2|}^2g_2^{ref}(\tau) \\
                                 & + (I_1{|\alpha _2|}^2+I_2{|\alpha _1|}^2)[1+\xi |\gamma (\tau)|\cos(\varphi _\gamma +\Delta \phi_\alpha )],
        \end{aligned}
    \end{equation}
where $\xi\equiv 2 |\alpha _1\alpha _2|\sqrt{I_1I_2}/(I_1{|\alpha _2|}^2+I_2{|\alpha _1|}^2)$ with $I_{1,2}$, $|\alpha_{1,2}|^2$ being the photon rates of the incoming fields $E_1, E_2$ and the reference fields, respectively, $\gamma (\tau)=|\gamma|e^{i\varphi_\gamma}$ is the second-order coherence function between $E_1, E_2$ that will be measured, and $g_2(\tau), g_2^{ref}(\tau)$ are the normalized intensity cross-correlation functions between $E_1, E_2$ and between two reference fields, respectively. In deriving Eq.(\ref{corr}), we assumed that the incoming fields $E_1, E_2$ and the reference fields are totally uncorrelated and that the reference fields are path-balanced, so that  they are completely coherent with $|\gamma_{12}^{ref}|=1$ and a definite phase difference $\Delta \phi_\alpha$. The last term is the interference term, corresponding to the signal, and can be made large by increasing the reference fields $|\alpha_{1, 2} |^2$, while the first two terms are the background noise, which is mostly dominated by the second term, since the incoming celestial light is usually very weak. So, the background also increases with reference fields. The overall consequence can be seen through the visibility of the interference, which is a measure of the signal size and has the following form when we set $I_1 = I_2\equiv I, |\alpha_1| =|\alpha_2|\equiv |\alpha|$:
    \begin{equation}\label{V}
    V= \frac{2I|\alpha|^2|\gamma_{12} |}{|\alpha|^4 g_2^{ref}+ 2I(|\alpha|^2+I)},
    \end{equation}
where we set $g_2=2$ for the thermal fields that are usually from celestial light. For coherent states as  reference fields \cite{tang25}, we have $g_2^{ref}=1$, and $V$ is significant only when $|\alpha|^2 \sim I$, as shown in the black curve in Fig.\ref{Vi} where we plot visibility $V$ as a function of $|\alpha|^2/I$ for various values of $g_2^{ref}$. But this leads to a coincidence measurement signal of $R_c \propto \Gamma_{2,2} \sim I^2$, which gives the same signal level as the HBT method, and is, for weak celestial light, much smaller than the traditional Michelson stellar interferometry, which is proportional to $I$. On the other hand, if $g_2^{ref}=0$, or the reference fields are ideally  anti-correlated such as the ones split from a field in single-photon state or anti-bunched source, we have $V\rightarrow 1$ for $|\alpha|^2/I\gg 1$ from Eq.(\ref{V}) and Fig.\ref{Vi} (red trace), so we can increase the signal rate without the reduction of visibility. This is the idea of Grottesman et al \cite{gott} of using a path-entangled single-photon state as the reference fields since such a single-photon state ideally has $g_2^{ref}=0$.

Furthermore, as shown in Fig.\ref{Vi}, even for  non-ideally anti-correlated reference fields with $g_2^{ref} \ne 0$ (blue curves), $V$ drops slower than coherent state references (black curve, $g_2^{ref} =1$). This means that we can increase $|\alpha|^2$ to enhance signal level without compromising visibility.

\section{Practical considerations}

In practice,  the anti-correlated reference fields can be produced by beam splitting a single-photon state or an anti-bunched optical field. However, we only have anti-bunched fields with $g_2^{ref}(\tau)=0$ for $\tau \lesssim T_c^{ref}$, the correlation time of the reference fields, and $g_2^{ref}(\tau)\rightarrow 1$ for $\tau \gtrsim T_c^{ref}$. So, there would be a high probability of finding two photons beyond $T_c^{ref}$ and the separation between two photons is $T_c^{ref}$ on average, or the photon rate $|\alpha|^2$ of the reference fields is limited by $1/T_c^{ref}$. 

Furthermore, we experimentally measure the coincidence count within a certain coincidence window of $T_R$, which is related to the intensity correlation function in Eq.(\ref{corr}) through the coincidence rate as  \cite{mandel-wolf}
    \begin{equation}\label{coinc}
            R_c = \int_{T_R} d\tau~\Gamma_{2,2}(\tau ) .
    \end{equation}
So, the signal rate is proportional to $T_R$, which must be smaller than $T_c$, the coherence time of the incoming field in order to resolve $\gamma(\tau)$ in time because an integration on $\gamma(\tau)$ is performed in Eq.(\ref{coinc}). $T_R$ is further limited by $T_c^{ref}$, the range in which $g_2^{ref}(\tau) \sim 0$. All this thus sets a limit on how large we can have the reference field strength $\alpha$ to enhance the signal level of coincidence measurement. 

Ideally, if we could have a source that emits a photon deterministically in a fixed interval of $T_c^{ref}$, then the anti-bunched photon source has a step function shape with $g_2^{ref}(\tau)=0$ for $\tau\le T_c^{ref}$ and $|\alpha|^2$ as large as $1/T_c^{ref}$. We can then set $T_R = T_c^{ref}$ so that we have from Eqs.(\ref{corr}) and (\ref{coinc}) 
    \begin{eqnarray}\label{coinc-id}
            R_c & \approx & 2 T_R I|\alpha|^2 [1+ |\gamma|\cos(\varphi _\gamma +\Delta \phi _\alpha )]\cr
            &=& 2I[1+ |\gamma|\cos(\varphi_\gamma +\Delta \phi _\alpha )],
    \end{eqnarray}
where we dropped the $I^2$ term assuming weak incoming fields. This is exactly the same signal rate as the traditional Michelson stellar interferometric method where two fields, each having a photon rate of $I$, interfere to measure the $\gamma$-function.

However, such an ideal source does not exist in the lab. Most of the common sources of anti-bunched photons \cite{tom21,ding25} are based on the resonance fluorescence process \cite{kim77}, where a coherent field on resonance with a single two-level emitter is incoherently scattered. They usually have $g_2^{ref}(\tau)\ne 0$ for $\tau\ne 0$. Let us examine what this type of photon source can offer us, and we will use the parameters of this photon source for optimization of the signal-to-noise ratio (SNR).

\section{Realistic single-photon state from anti-bunched photon sources of resonance fluorescence}

 A common method to produce the spatially entangled single-photon state for the GJC quantum telecopic scheme is to split a single-photon state or an anti-bunched photon source by a 50:50 beam splitter, as shown in the red dashed box in Fig.\ref{HBT}(b). In this section, we consider a realistic anti-bunched photon source from the atomic resonance fluorescence process where a single two-level atom is continuously driven by a coherent excitation source such as a cw laser and undergoes cyclic excitation and emission processes. At the moment right after the emission of one photon, the single atom falls back to the ground state and the probability of emitting a subsequent photon is non-zero only at some delayed time after the prior emission event when the atom is re-pumped to the excited state.

\begin{figure}[t]
        \includegraphics[width=8cm]{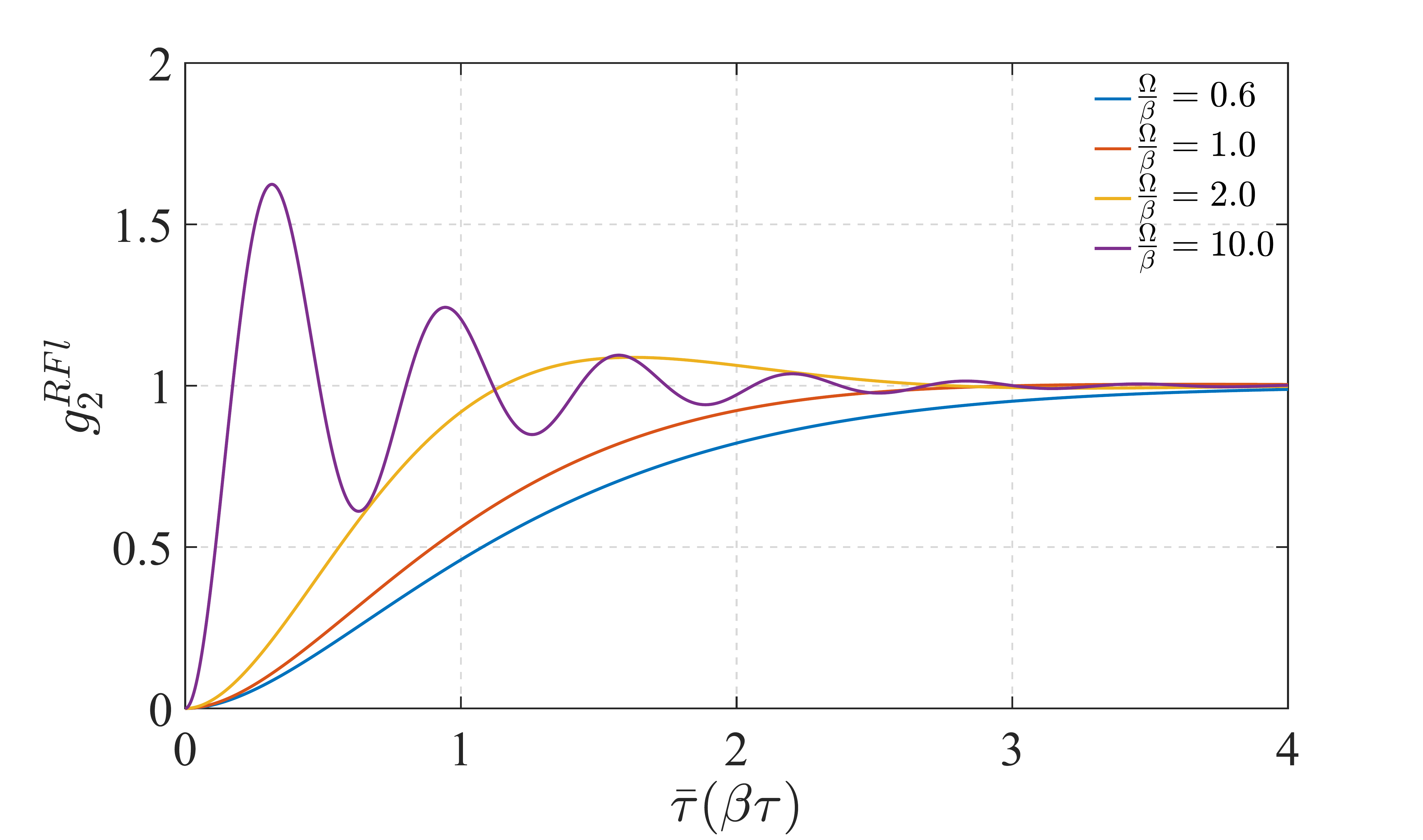}
            \caption{Normalized intensity correlation function $g^{RFl}_2$ of resonance fluorescence verses $\bar\tau \equiv \beta\tau$ for $\Omega/\beta=$10, 2, 1, 0.5.} 
            \label{g2}
    \end{figure}

The theory of resonance fluorescence (RFl) was worked out by Kimble and Mandel in 1976 \cite{kim76}, where 
the normalized intensity correlation function $g_2^{RFl}(\tau)$ of the scattered fluorescence was calculated, showing the signature anti-bunching effect \cite{kim77,dag78}. In general, the $g_2$-function depends on the Rabi frequency $\Omega\equiv E_0d/\hbar$ ($d$ is the atomic transition dipole moment and $E_0$ is the peak amplitude of the electric field of the coherent driving field), the decay constant $\beta$ of the upper atomic energy level or half the Einstein A-coefficient, and the detuning $\Delta\equiv \omega_L-\omega_a$ of the driving frequency $\omega_0$ from the atomic transition frequency $\omega_a$. The exact form is quite complicated and is presented in Eq.(\ref{g-RFl}) in Appendix A for zero detuning. To see the behavior of $g_2^{RFl}(\tau)$ in Eq.(\ref{g-RFl}), we plot it in Fig.\ref{g2} for four values of $\Omega/\beta (0.5, 1, 2, 10)$ as a function of the dimensionless delay time $\bar \tau \equiv \beta\tau$. The time correlation function $g_2^{RFl}(\bar \tau)$ begins with a value of zero at $\bar \tau = 0$ immediately after a photon is emitted, but rises for non-zero $\bar \tau$ and asymptotically approaches unity for large $\bar \tau$ through a series of damped oscillations. The number of these oscillations depends primarily on the pump strength $\Omega/\beta$. A higher $\Omega/\beta$ results in a steeper rising slope, which shortens the interval between successive photon emissions. As can be seen, $g_2^{RFl}(\tau)$ does not have the ideal step function shape required to obtain Eq.(\ref{coinc-id}) in the last section.

The photon rate of the scattered light is quite simple and is given by
    \begin{equation}\label{alph}
       |\alpha|^2 = \beta \frac{\frac{1}{2}\Omega^2/\beta^2 }{\frac{1}{2}\Omega ^2/\beta ^2+1} \rightarrow \beta~~~~{\rm for~~\Omega \gg \beta}.
    \end{equation}
Therefore, when pumped hard ($\Omega \gg \beta$), the photon rate and the correlation time $T_c^{RFl}$ are determined by $\beta$ and the Rabi frequency $\Omega$, respectively.

So, given the level $I$ of the incoming light and its bandwidth $\Delta \omega = 1/T_c$, we need to optimize the detector parameter of $T_R$ and the source parameters of $\beta, \Omega$ of the anti-bunched reference fields to maximize the measurement output. 

\section{Signal-to-Noise Ratio (SNR)}

Although the visibility of interference provides a sense of the signal size against the background, the metric that truly matters in a measurement is the signal-to-noise ratio. We can derive it from the coincidence count rate $R_c$ through Eqs.(\ref{corr},\ref{coinc}). In the experiment, we measure the total coincidence count $N_c = R_cT$ for a data accumulation period $T$. The ``Signal'' part is the size of the interference fringe, and from Eq.(\ref{corr}) with $I_1=I_2=I, \alpha _1=\alpha _2=\alpha$ for maximum visibility, we find it as the coefficient in front of the cosine function:
    \begin{equation}\label{sig}
            {\rm Signal} = R_c^{signal} T= 2 T I{|\alpha |}^2\int_{T_{R}} d\tau |\gamma (\tau)|.
    \end{equation}
The ``Noise'' part is the fluctuation of the total coincidence count $N_c$. Since the data accumulation time $T$ is quite long for weak incoming light, it is usually much longer than the correlation time of all fields. So, the coincidence events are all uncorrelated and can be treated as a Poissonian process for which the fluctuation is $\Delta N_c = \sqrt{N_c}$. But $N_c$ changes in the data set due to interference (the cosine term).  Since we usually take the data of a full fringe to measure the ``Signal'' part, we use an average over one full fringe for evaluating ``Noise'',  which just removes the cosine term and keeps the baseline terms:
            \begin{eqnarray}\label{noise}
                    {\rm Noise} &=&\sqrt{N_c^{ave}} \cr
                  &=& \sqrt{T\int_{T_{R}} d\tau \left \{ I^2 g_2 (\tau)+|\alpha |^4 g_2^{ref}(\tau)+2I{|\alpha |}^2 \right \} }.\cr &&
            \end{eqnarray}
Notice that we do not include dark counts in our analysis. They will contribute as accidental coincidences and raise the baseline to add extra noise.

Combining Eq.(\ref{sig},\ref{noise}), we find the SNR as follows:
    \begin{eqnarray}    \label{SNR-g}
                 SNR &=& \frac{2I\sqrt{T}\int_{T_{R}} d\tau |\gamma (\tau)|}{\sqrt{\int_{T_{R}} d\tau \left \{ \frac{I^2}{|\alpha |^4} g_2(\tau)+g_2^{ref}(\tau)+\frac{2I}{|\alpha |^2} \right \} }} \cr
                 &=&  SNR_{MI}\frac{\sqrt{2N_R}~\bar\gamma(T_R)}{\sqrt{\frac{I^2}{|\alpha |^4} \bar g_{2}(T_R)+\bar g_{2R}^{ref}+\frac{2I}{|\alpha |^2} }},        
    \end{eqnarray}
where $SNR_{MI}\equiv \sqrt{2I T}$ is the SNR for Michelson stellar interferometry, $N_R\equiv I T_R$ is the photon number of the incoming field within the coincidence window $T_R$.  $\bar{\gamma}(T_R)\equiv \frac{1}{T_R}\int_{T_R} d\tau |\gamma (\tau)|$, $\bar g_2(T_R)\equiv \frac{1}{T_R}\int_{T_R} d\tau g_2(\tau)$, and $\bar g_{2R}^{ref}\equiv \frac{1}{T_R}\int_{T_R} d\tau g_2^{ref}(\tau)$ are the averages of the corresponding quantities within the coincidence window $T_R$ and are close to 1, $g_2(0)$, and  $g_2^{ref}(0)$, respectively, for $T_R\ll T_c$, the coherence time of the incoming field (This condition is for time-resolved coincidence measurement of the complex function $\gamma (\tau)$). 

    \begin{figure*}[t]
        \includegraphics[width=8cm]{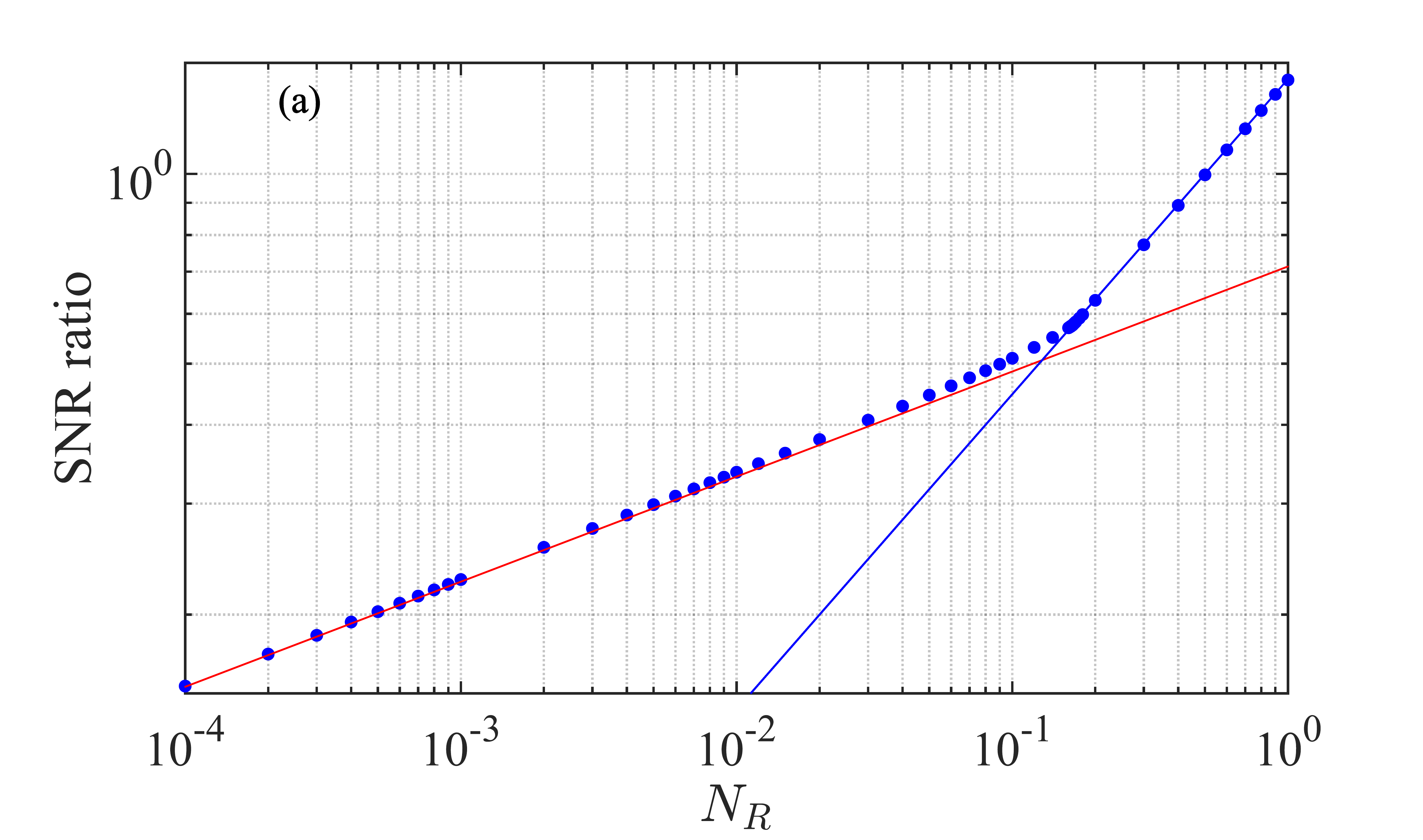}
        \includegraphics[width=8cm]{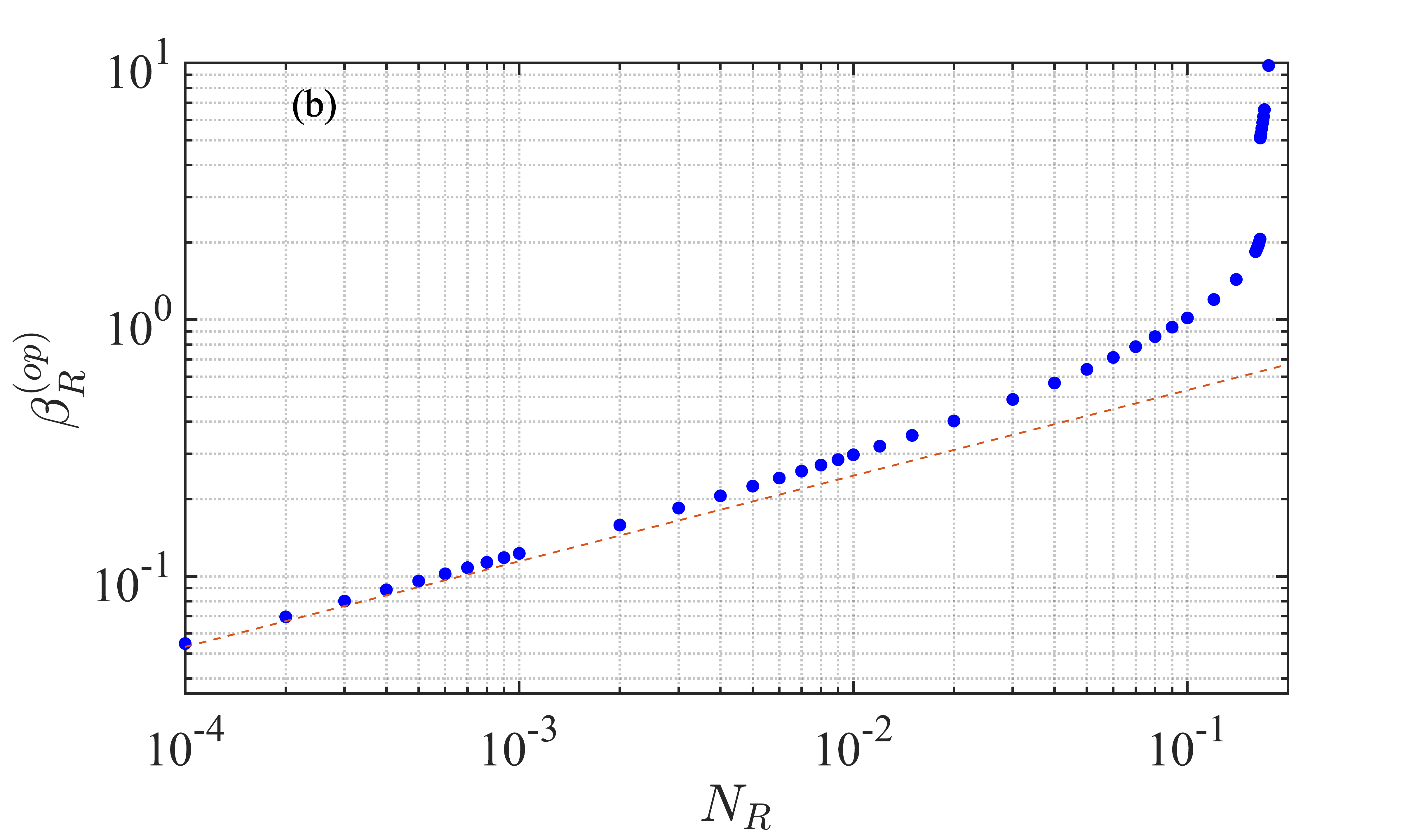}
            \caption{(a) The optimized SNR ratio and (b) $\beta_R^{(op)}$  versus the photon number $N_R$ of the incoming field. Circles are from numerical simulation. The red line in (a) is from Eq.(\ref{op}) for the scheme with resonance fluorescence and the blue line in (a) is from Eq.(\ref{snr-coh}) for the coherent state scheme. The dashed red line in (b) is from Eq.(\ref{op}) for $\beta_R^{(op)}$.}
            \label{fig-ratio}
    \end{figure*}
    
For coherent states as reference fields with $g_2^{ref} = 1$, we find that the SNR is maximum when the coherent reference fields are much stronger than the incoming fields: $|\alpha|^2 \gg I$ with
    \begin{eqnarray}\label{snr-coh}
                 SNR_{coh} = SNR_{MI}\sqrt{2N_R}.       
    \end{eqnarray}
So, for weak celestial light with $N_R\ll 1$, the phase-sensitive HBT intensity stellar interferometry has a much smaller SNR than the Michelson stellar interferometry. 

Notice that from Eq.(\ref{SNR-g}), we find that by using anti-bunched photon sources even with $0 \ne g_2^{ref}<1$, we can still enhance the SNR by a factor of $1/\sqrt{g_2^{ref}}$ when $I/|\alpha|^2\ll g_2^{ref} < 1$:
    \begin{eqnarray}
                 SNR_{anti} = SNR_{coh}/\sqrt{g_2^{ref}}.       
    \end{eqnarray}

Next, we will consider anti-bunched photon sources from resonance fluorescence (with $g_2(0)=0$) as the reference fields and find how it compares with both the coherent state scheme and the Michelson scheme.

    \section{Optimization of SNR for resonance fluorescence as reference sources}

    Now let us examine the anti-bunched photon source of resonance fluorescence as the reference field, which has the adjustable parameters of $\beta, \Omega/\beta$. 
    For perfect alignment in the resonance fluorescence collection system, we have the mean photon rate of the source given in Eq.(\ref{alph}). Substituting it into Eq.(\ref{SNR-g}), we have the SNR for resonance fluorescence (RFl):
        \begin{eqnarray}\label{finalSNR}
          && \frac{SNR_{RFl}}{SNR_{MI}} \cr &&\hskip 0.3in =\frac{\sqrt{N_R}}
            {\sqrt{ \frac{N_R^2}{\beta_R^2}\big(\frac{\Omega ^2+2\beta ^2}{\Omega ^2} \big)^2 + \frac{1}{2}\bar g_{2R}^{RFl}(\beta_R,\frac{\Omega}{\beta})+\frac{N_R}{\beta_R}\frac{\Omega ^2+2\beta ^2}{\Omega ^2}}} , \cr &&
        \end{eqnarray}
where we set $\bar{g}_2 (T_R)\approx 2, \bar\gamma (T_R)\approx 1$ by assuming that the incoming field is thermal light and $T_R\ll T_c$.  $\beta_R\equiv \beta T_R$ and $\bar g_{2R}^{RFl}(\beta_R,\Omega/\beta)\equiv \frac{1}{\beta_R}\int_{\beta_R} d\bar \tau g_2^{RFl}(\bar \tau,\Omega/\beta)$ with $\bar \tau\equiv \beta \tau$, and $g_2^{RFl}(\bar \tau,\Omega/\beta)$ given in Eq.(\ref{g-RFl}) for resonance fluorescence.

With fixed incoming light level $I$ and bandwidth $\Delta \omega = 1/T_c$, we choose the detector response with fixed $T_R = (0.02\sim 0.2)T_c$ for time-resolved coincidence measurement. So, $N_R=I T_R$ is fixed. We need to optimize $\beta_R, \Omega/\beta$ for the best $SNR_{RFl}$. 

Due to the complicated form of $g_2^{RFl}$, the optimization process has to be performed numerically and is quite tedious. We present it in Appendix B. However, for $N_R\ll 1$, we can obtain a simple analytical result summarized as 
        \begin{eqnarray}\label{op}
            &&(\Omega/\beta)^{(op)} = 2~~ {\rm independent~of}~N_R,\cr
            && \beta_R^{(op)} = {\left(3N_R/2 \right)}^{\frac{1}{3}},\cr
            && SNR_{RFl}^{(op)} / SNR_{MI} = {\left(2/3 \right)}^{\frac{5}{6}} {N_R}^{\frac{1}{6}} .
        \end{eqnarray}

For $N_R\sim 1$, we resort to numerical evaluation and optimization. The process is quite complicated and is presented in the Supplementary Materials \cite{sup}. We only display the results here. 
The numerically optimized SNR ratio $SNR_{RFl}^{(op)}/SNR_{MI}$ and $\beta_R^{(op)}$ (circles) are plotted versus the photon number $N_R$ of the incoming field in Fig.\ref{fig-ratio}(a,b) on a log-log scale. As can be seen, the dependence expressed in Eq.(\ref{op}) is followed quite well for small $N_R$ (the red solid line in (a) and the red dashed line in (b) are from Eq.(\ref{op})). However, after $N_R$ reaches 0.16, $\beta_R^{(op)}$ begins to increase drastically and $SNR_{RFl}^{(op)} \rightarrow SNR_{coh}$ (The blue line in (a) is from Eq.(\ref{snr-coh})).

\begin{figure}[t]
        \includegraphics[width=8cm]{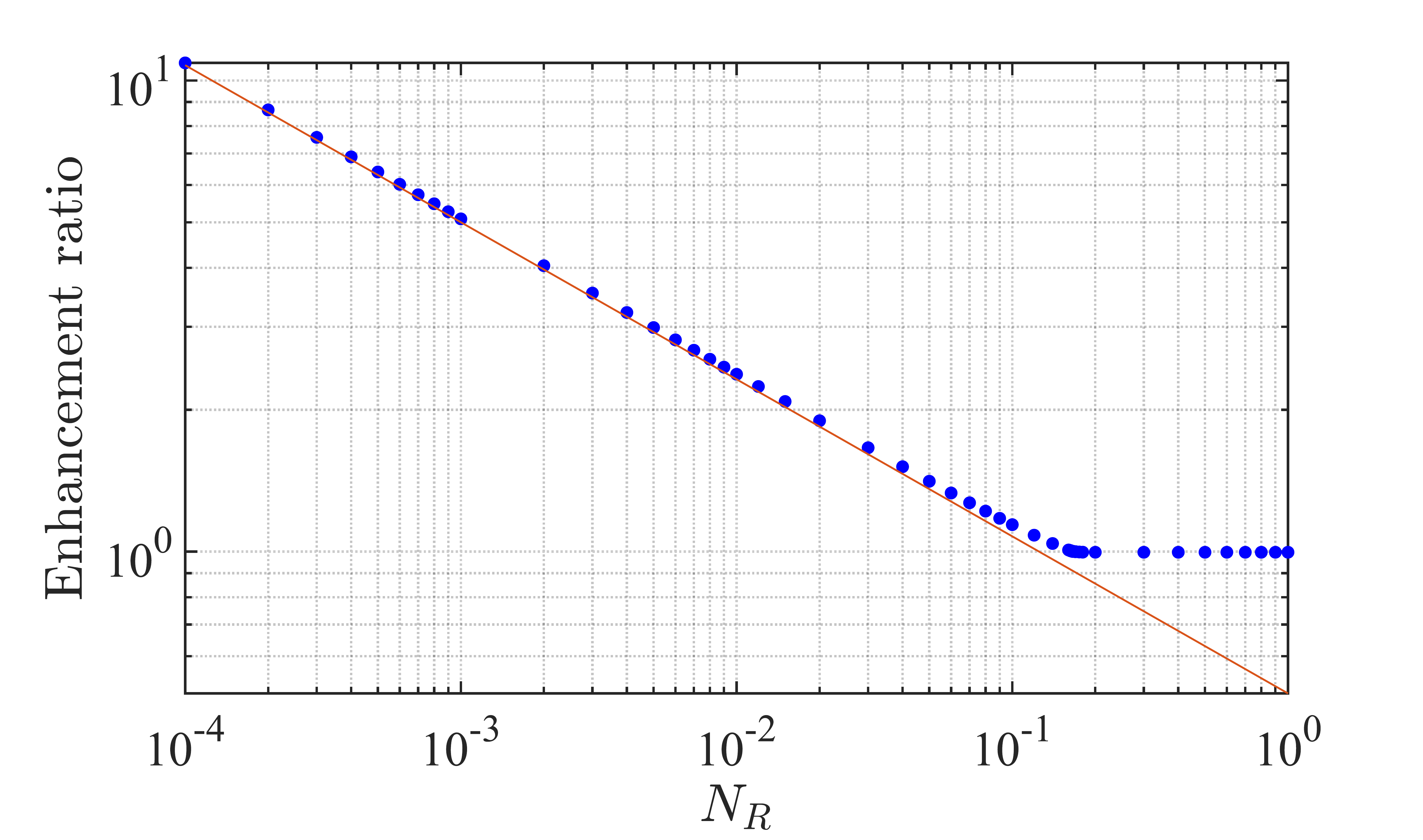}
            \caption{Enhancement ratio as a function of input photon number $N_R$ in log scale. The solid line is from Eq.(\ref{ratio-coh}). }
            \label{Enhance_I}
    \end{figure}

For small $N_R$, using an anti-bunched photon source as reference fields can indeed improve SNR as compared to the scheme with the coherent state as reference. This can be seen by comparing the result in Eq.(\ref{op}) with the result in Eq.(\ref{snr-coh}) for the scheme of coherent state as reference fields and obtain an enhancement ratio as
        \begin{eqnarray}\label{ratio-coh}
 SNR_{RFl}^{(op)} / SNR_{coh} \approx 0.5 {N_R}^{-1/3} .
         \end{eqnarray}
The numerically calculated enhancement ratio (circles) is plotted against $N_R$ in Fig.\ref{Enhance_I}, which shows a higher enhancement for weaker input ($N_R\ll1$) but no enhancement when $N_R \rightarrow 1$. The solid line is from Eq.(\ref{ratio-coh}), which fits well with the circles for small $N_R\ll1$.

From Fig.\ref{fig-ratio}(a), we find that for a weak incoming field ($N_R\ll 1$), the SNR with resonance fluorescence as reference fields is smaller than that of the Michelson scheme, and the SNR ratio decreases with the drop of $N_R$. This is due to the non-ideal shape of $g_2^{RFl}$ as shown in Fig.\ref{g2}, where $g_2^{RFl}$ arises from zero rather quickly. 

\begin{figure}[t]
        \includegraphics[width=8cm]{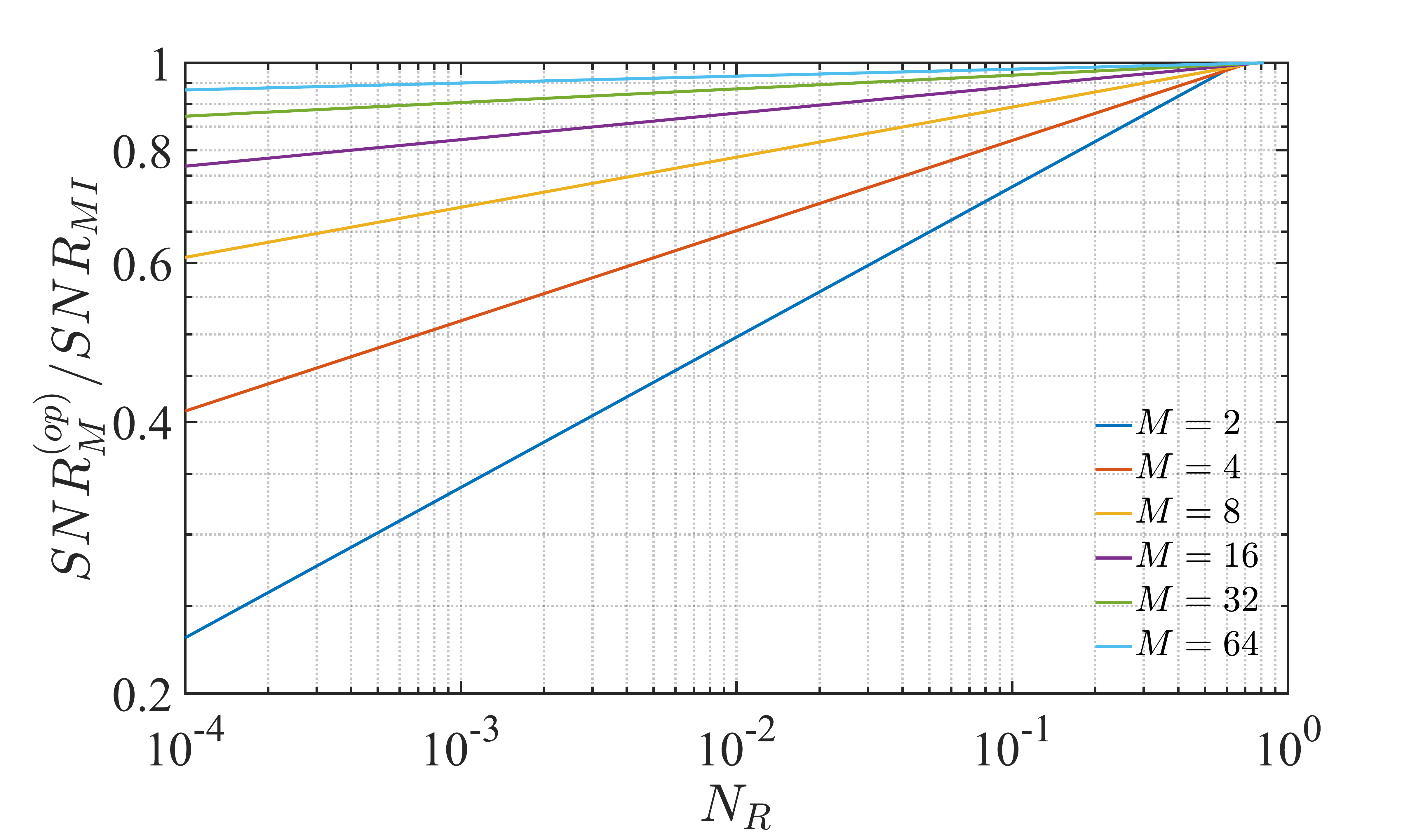}
            \caption{Optimized SNR ratio $SNR_M^{(op)}/SNR_{MI}$ given in Eq.(\ref{op2}) as a function of $N_R$ for various values of $M$}
            \label{M}
    \end{figure}

In deriving Eq.(\ref{coinc-id}), we used an unrealistic step function shaped $g_2^{ref}(\tau)$ function with a completely flat bottom at zero. A realistic source of resonance fluorescence gives a $g_2^{RFl}$ arising from zero in the $\tau^2$-dependence (see Appendix B). Let us now consider a flatter bottom of $g_2^{ref}$ than $g_2^{RFl}$ with a slowly increasing value to approximate the completely flat bottom.  A high power dependence of $g_2^{ref} \propto \tau^M (M>2)$ for $\tau$ near zero has a slower increasing rate than the $\tau^2$-dependence of $g_2^{RFl}$ near $\tau=0$. We work out this in Appendix B for the general case in Eq.(\ref{SNR-g}) with $N_R\ll 1$ and obtain the following results:
    \begin{eqnarray}    \label{SNR-g3}
                 SNR_M^{(op)} 
                 &=& SNR_{MI}\left(\frac{M}{M+1}\right)^{\frac{M}{2(M+1)}}\left(\frac{CN_R}{2}\right)^{\frac{1}{2(M+1)}} \cr
                 &\propto & SNR_{MI} \times N_R^{\frac{1}{2(M+1)}},
    \end{eqnarray}
with the optimized value achieved when the photon number of the reference field  $N_R^{ref} \equiv |\alpha|^2T_R= [2N_R(M+1)/CM]^{\frac{1}{M+1}}$, where we assume $g_2^{ref}(\tau) \approx C(|\alpha|^2\tau)^M (M>2)$ for $|\alpha|^2\tau\ll 1$ with $C$ as some dimensionless quantity of size around 1.  Figure \ref{M} plots the optimized SNR ratio $SNR_M^{(op)}/SNR_{MI}$ as a function of $N_R$ for various values of $M$ ($C$ is set as 27/4, the value for resonance fluorescence). So, as the bottom of $g_2^{ref}$ becomes flatter ($M\rightarrow$ large), the SNR approaches $SNR_{MI}$. 

The bottom-flattened $g_2^{ref}$ can be engineered by the spectral structure of the anti-bunched photon source.

\section{EFFECT OF LOSS IN REFERENCE}

Until now, our analysis has focused on the ideal scenario with no loss in reference fields. However, in any realistic implementation of a single-photon state, loss is inevitable due to such effects as inefficient single-photon collection, absorption and scattering in propagation, and mode mismatch in interference.

The effect of loss $L$ on the reference field from resonance fluorescence is simply to add a factor of $1-L$ in front of Eq.(\ref{alph}):
    \begin{equation}\label{alph-L}
       |\alpha_L|^2 = (1-L) \beta \frac{\frac{1}{2}\Omega^2/\beta^2 }{\frac{1}{2}\Omega ^2/\beta ^2+1} .
    \end{equation}
The optimization process is still the same, but the results for $N_R\ll 1$ become
        \begin{eqnarray}\label{op2}
            (\Omega/\beta)^{(op-L)} &=& 2~~ {\rm independent~of}~N_R,\cr
             \beta_R^{(op-L)} &=& {\left( \frac{3N_R/2}{1-L} \right)}^{1/3},\cr
            \frac{SNR_{RFl}^{(op-L)}}{SNR_{MI}} &=& (1-L)^{1/3}{\left(\frac{2}{3} \right)}^{\frac{5}{6}} {N_R}^{1/6} .~~~~
        \end{eqnarray}
So, the SNR drops by a factor of $(1-L)^{1/3}$.

For the case of a flatter bottom of $g_2^{ref}$ with $g_2^{ref}\propto \tau^M$ for $\tau\rightarrow 0$, we need to replace $|\alpha|^2$ in Eq.(\ref{SNR-g}) with $(1-L)|\alpha|^2$.  For $N_R\ll 1$, it is straightforward to find (through Eq.(\ref{SNR-g2}) in Appendix B)
        \begin{eqnarray}\label{op4}
        \frac{SNR_M^{(op-L)}}{SNR_{MI}} = (1-L)^{\frac{M}{2(M+1)}}\frac{SNR_M^{(op)}}{SNR_{MI}}.
        \end{eqnarray}
So, the drop factor $= (1-L)^{M/2(M+1)} \rightarrow (1-L)^{1/2} (M\rightarrow \infty)$ is slightly more for larger $M$ (flatter bottom).

However, for $N_R\sim 1$, we find from Eq.(\ref{SNR-g}) after the replacement of $|\alpha|^2$ by $(1-L)|\alpha|^2$, it is equivalent to the change of $N_R$ to $N_R/(1-L)$ with a significant increase for the large loss of $1-L\ll 1$. According to Fig.\ref{Enhance_I}, this may lead to the drop in the enhancement effect over the coherent state scheme and eventually to no enhancement. So, a large loss of single-photon states makes it equivalent to a weak coherent state. 

It should be mentioned that the spatially entangled single-photon state used here is usually produced by beam splitting a single-photon state and its photon rate is therefore reduced by half or $L=0.5$ \cite{tsa11,zhang2025}. So, $SNR_{RFl}$  is reduced by $(1-L)^{1/3} = 1/\sqrt[3]{2}$ according to Eq.(\ref{op2}).

    \section{Conclusion and discussion}

Although numerous challenges remain in implementing very-long-baseline telescopes with single-photon states as references, we identify in this work another issue related to the temporal correlation of the single-photon state. Through theoretical and numerical analysis, we demonstrate that an ideal single-photon reference exhibiting a step-like correlation function is preferred. This will minimize the background coincidence counts from the reference while maintaining a high intensity for the reference fields in order to increase the signal level. Under ideal conditions, the signal level achievable with a perfect single-photon state becomes comparable to that of the direct interference method of Michelson stellar interferometry. 

However, practical single-photon sources, such as those from resonance fluorescence, have a shape of temporal correlation far from ideal, leading to a signal-to-noise ratio (SNR) lower than that of the Michelson method, especially at low incoming photon number. From our analysis, we find that spectral engineering to change the shape of the temporal correlation can improve the SNR, eventually matching that of the Michelson method.  On the other hand, using anti-bunched photon sources is still better than using coherent states as the reference field, providing quantum enhancement to SNR with the enhancement factor increasing for a lower input photon number. 

In our investigation, the quantity $N_R \equiv I T_R$ appears constantly. This is the photon number of the incoming field within the detector's resolution time $T_R$. To see its meaning, we can compare it to a quantity commonly known as the occupation number $\delta$ per mode. For CW fields, the coherence time $T_c$ of the fields is known to define a single temporal mode.  So, $\delta = I T_c$ is the average photon number per mode, or the so-called occupation number per mode. Since we require $T_R< T_c$ for time-resolved coincidence measurement, $N_R < \delta$ is then the average photon number per mode measured by the detector. $\delta = 1/(e^{h\nu/kT} -1)$ for a blackbody of temperature $T$. The Sun has $\delta \approx 0.01$ at 555 nm but in the microwave background radiation such as the cosmic microwave background, $\delta \approx 10^8$.  We can estimate $N_R$ for some celestial sources. For example, for the brightest star Sirius, from the spectral data  $F_{\lambda} = 1.35\times 10^{-8} erg~ cm^{-2} {\AA}^{-1}s^{-1}$ at 555 nm \cite{hubble}, we obtain a photon flux of $I= 9.5 \times 10^7/s$ per polarization mode for an area of 1 $m^2$ area and a spectral width of 0.5 nm (coherence time $T_c$ =20 ps). Taking $T_R = T_c/5 = 4$ ps for the time resolved measurement, we arrive at $N_R= IT_R \sim 4 \times 10^{-4} \ll 1$. This photon level justifies the use of the single-photon state as reference fields for the enhancement of the signal level. 

The condition $N_R\ll 1$ in our analysis means that the average photon number per detected mode is much smaller than 1, corresponding to weak incoming fields. Our analysis shows that when $N_R\sim 1$, there is no advantage in using an entangled single-photon state as the reference fields compared to the coherent state. This is consistent with the single-mode analysis performed by Tsang \cite{tsa11} and is also in line with the SNR analysis of the homodyne/heterodyne detection technique \cite{homo1,homo2}, where the shot noise is the limiting factor for weak incoming fields \cite{homo3}. The large occupation number in the microwave explains well why coherent states are good enough as LO for radio telescope arrays.

Since phase-dependent Hanbury Brown–Twiss (HBT) interferometry is based on interference between all fields, spatial mode matching is essential for achieving high visibility. Our current analysis focuses primarily on temporal mode match through time resolved coincidence measurement (single-mode). The spatial mode mismatch will reduce the visibility of interference, and thus the signal size. It is equivalent to the loss effect in detection. We did not consider inefficient detectors in our analysis. Since coincidence between two detectors is measured, the signal size will drop by a factor of $\eta^2$ with $\eta$ as the quantum efficiency of the detectors. This is a disadvantage compared to the drop factor of $\eta$ for the direct intensity measurement by one detector in the Michelson method.

\acknowledgments
The authors thank Professor M. Raymer for challenging them to consider the issue of signal-to-noise ratio in various schemes in this paper. The work was supported by Joint NSFC/RGC Collaborative Research
Scheme (CRS-CityU103/24), the General Research Fund from Hong
Kong Research Grants Council (Grant No. 11301624).

\bibliography{references}

\vskip 0.1in
\appendix
\section{$g_2$-function from resonance fluorescence}

The theory of resonance fluorescence was worked out by Kimble and Mandel in 1976 \cite{kim76}. The results can be written as follows. For simplicity, we set the detuning to zero, then
the normalized intensity correlation function $g_2^{RFl}(\tau)$ is given by

\begin{widetext}
            \begin{equation}\label{g-RFl}
                \begin{aligned}
                    g_2^{RFl}(\tau) &= 1 - \left(\frac{A+\frac{1}{3}}{B^{2}+9 A^{2}}\right)\frac{\left(\frac{1}{9}-\frac{4}{3} A+4A^{2}\right)\left(\frac{1}{2}\Omega^{2}/\beta^{2}+1\right)}{\left(\Omega^{2}/\beta^{2}\right)\left(\frac{1}{2} A-\frac{1}{3}\right)}e^{(-4/3+2A)\beta\tau}  \\
                    &\quad - \frac{4\left(\frac{1}{2}\Omega^{2}/\beta^{2}+1\right)}{\Omega^{2}/\beta^{2}} e^{-(4/3+A)\beta\tau}\left(\frac{FH+GJ}{H^{2}+J^{2}}\cos B\beta\tau-\frac{GH-FJ}{H^{2}+J^{2}}\sin B\beta\tau\right)
                \end{aligned}
            \end{equation}
\end{widetext}
where
    $A = \frac{1}{2}\left(\eta_{+} + \eta_{-}\right), ~~
    B = \frac{1}{2}\sqrt{3}\left(\eta_{+} - \eta_{-}\right) $
with
\begin{eqnarray}
    \eta_{\pm} &\equiv& \frac{1}{3}\bigg\{
        9\Omega^{2}/2\beta^{2} - 1
        \pm \Big[
            \Big(9\Omega^{2}/2\beta^{2} - 1\Big)^{2}
             \cr
        &&\qquad\qquad +\left(3\Omega^{2}/\beta^{2} - 1\right)^{3}
        \Big]^{1/2} \bigg\}^{1/3},
\end{eqnarray}
and
\begin{eqnarray}
    F &=& \frac{2}{27} + \frac{1}{3}A - A^{3} + 3AB^{2}, \\
    G &=& -\frac{1}{3}B - B^{3} + 3A^{2}B, \\
    H &=& 8AB^{2} + \frac{8}{3}B^{2}, \\
    J &=& 8AB + 6A^{2}B - 2B^{3},
\end{eqnarray}
which is a function of $\Omega/\beta$ and $\beta\tau$ with $\Omega$ as the Rabi frequency and $\beta$ as the half Einstein A-coefficient. Figure \ref{g2} shows $ g_2^{RFl}$ as a function $\bar \tau\equiv \beta\tau$  for various values of $\Omega/\beta$. 

\section{Optimization of the SNR}

For fixed $N_R$, we just need to minimize the expression inside the square root in the denominator of Eq.(\ref{finalSNR}):
        \begin{eqnarray}\label{denom}
            D &\equiv &
            \left(\frac{N_R}{\beta_R} \frac{\frac{1}{2}\Omega^2/\beta ^2+1}{\frac{1}{2}\Omega ^2/\beta ^2} \right)^2 + \frac{1}{2}\bar g_{2R}^{RFl}(\beta_R,\Omega/\beta)\cr &&\hskip 0.5in +\frac{N_R}{\beta_R} \frac{\frac{1}{2}\Omega^2/\beta^2+1}{\frac{1}{2}\Omega ^2/\beta ^2} .
        \end{eqnarray}

Notice that for celestial light, $N_R$ is usually quite small. For example, the brightest star of Sirius has a spectral flux density of $1.35 \times 10^{-11}W/cm^2/\mu m$ \cite{hubble}. For a 1 meter size telescope of 1 GHz bandwidth ($T_c=1 ns$), this translates to $I= 3\times 10^5$ photon/s or $N_R = 3\times 10^{-5} \ll 1$ for $T_R \sim 0.1 ns$ of a typical SNSPD in current photon detector technology. Therefore, we can neglect the first term in Eq.(\ref{denom}). Furthermore, we need the second term of $\bar g_2^{RFl}$ to be as small as possible, which means that the integral range of $g_2^{RFl}$ should be small so that $g_2^{RFl}$ is close to zero. In other words, $\bar \tau = \beta\tau$ should be much smaller than 1.

For $\bar \tau = \beta\tau \ll 1$, we can find the asymptotic expansion for $g_2^{RFl}(\bar \tau)$ in Eq.(\ref{g-RFl}). With the coefficients of the two exponential functions and the cosine and sine functions denoted as $T_1, T_2, C, S$, respectively, the asymptotic expansion has the following form
        \begin{eqnarray}\label{g-asym}
            g_2^{Rfl}(\tau) &=&1 - T_1 e^{(-4/3+2A)\beta\tau} - T_2 e^{-(4/3+A)\beta\tau} \cr
            &&\hskip 0.1in \times \bigl[ C\cos(B\beta\tau) - S\sin(B\beta\tau) \bigr] \nonumber\\
            &=& a_0+a_1\beta\tau+a_2\beta^2\tau^2+O\big((\beta\tau)^3\big) , 
        \end{eqnarray}
where
        \begin{eqnarray}
            a_0 &=& 1 - T_1 - T_2 C,\\[4pt]
            a_1 &=& -T_1\left(-\frac{4}{3} + 2A\right) + T_2 C\left(\frac{4}{3} + A\right) + T_2 S B,~~~~~~\\[4pt]
            a_2 &=& -\frac{T_1}{2}\left(-\frac{4}{3} + 2A\right)^2 
            - \frac{T_2 C}{2}\left(\frac{4}{3} + A\right)^2 \cr &&
            - T_2 S B\left(\frac{4}{3} + A\right) 
            + \frac{T_2 C}{2}B^2.
        \end{eqnarray}
After some lengthy derivation, it is straightforward to show: $a_0=0=a_1$, $a_2 = 1+ \frac{\Omega^2}{2\beta^2} $. With these, the expression in Eq.(\ref{denom}) becomes
        \begin{eqnarray}\label{denom2}
            D =
            \frac{\beta_R^2}{6}\left(1+\frac{\Omega^2}{2\beta^2}\right)+\frac{N_R}{\beta_R} \left(1+\frac{2\beta^2}{\Omega^2}\right), 
        \end{eqnarray}
which has a minimum value of 
\begin{eqnarray}\label{denom-m}
            D_{min} = \left(\frac{3}{2}\right)^{5/3} N_R^{2/3}
        \end{eqnarray}
when $\frac{\Omega}{\beta}=2$ and $\beta_R = \sqrt[3]{3N_R/2}$. This leads to a simple analytical optimization result in Eq.(\ref{op}) after substituting the expression inside the square root in Eq.(\ref{finalSNR}) with the minimum value above. 

As discussed in deriving Eq.(\ref{coinc-id}), a flat bottom of $g_2^{RFl}$ with a slowly increasing value is preferred. To see this quantitatively, we rewrite Eq.(\ref{SNR-g}) as
    \begin{eqnarray}    \label{SNR-g2}
                 SNR &=&  SNR_{MI}\frac{\sqrt{2N_R}}{\sqrt{\bar g_{2R}^{ref}+\frac{2N_R}{N_R^{ref}} }},        
    \end{eqnarray}
where we dropped the first term in the denominator and set $|\gamma|\approx 1$ due to $I/|\alpha|^2\ll 1$ and $T_R\ll T_c$, and $N_R^{ref}\equiv |\alpha|^2T_R$ is the average photon number in $T_R$ for the reference fields (Ideally, $N_R^{ref} = 1$ but usually $N_R^{ref} \lesssim 1)$. To have an estimate of  $\bar g_{2R}^{ref}$,  we rewrite it as 
    \begin{eqnarray}    \label{g2-ref}
                 \bar g_{2R}^{ref} &=&\frac{1}{T_R} \int_{T_R}d\tau g_{2}^{ref}(\tau)    \cr
                 &=& \frac{1}{N_R^{ref}} \int_{N_R^{ref}}d\bar \tau g_{2}^{ref}(\bar \tau)
    \end{eqnarray}
by changing the integral variable to dimensionless $\bar \tau \equiv |\alpha|^2\tau$. For an anti-bunched photon field with $g_{2}^{ref}(0)=0$ and a flat bottom of $g_{2}^{ref}(\bar\tau)$ with a slowly rising value, we assume that $g_{2}^{ref}(\bar \tau)$ has an asymptotic form of $g_{2}^{ref}(\bar \tau) \sim C\bar\tau^M$ for $\bar \tau\ll 1$ ($C\sim 1$ and $M$ are constant. $M=2$ for resonance fluorescence, see Appendix B). Then, Eq.(\ref{SNR-g2}) becomes
    \begin{eqnarray}    \label{SNR-g3}
                 \frac{SNR}{SNR_{MI}} &=&  \frac{\sqrt{N_R}}{\sqrt{\frac{C/2}{M+1}(N_R^{ref})^M+\frac{N_R}{N_R^{ref}} }}\cr
                 &\le& \left(\frac{M}{M+1}\right)^{\frac{M}{2(M+1)}}(CN_R/2)^{\frac{1}{2(M+1)}} ,~~~~~~
    \end{eqnarray}
with the maximum achieved when $N_R^{ref} = [2N_R(M+1)/CM]^{\frac{1}{M+1}}$. So, as the bottom of $g_2^{ref}$ becomes flatter ($M\rightarrow$ large), the SNR approaches $SNR_{MI}$. 

When $N_R\sim 1$, we cannot use the asymptotic form for $g_2^{RFl}$. The numerical evaluation is performed, but the procedure is rather complicated, so we place it in the Supplementary Materials \cite{sup} and only present the results in the main text as Figs.(\ref{fig-ratio}, \ref{Enhance_I}).

\end{document}